\documentclass[sigconf]{acmart}
\AtBeginDocument{%
  }

\setcopyright{acmlicensed}
\copyrightyear{2026}
\acmYear{2026}
\acmDOI{XXXXXXX.XXXXXXX}
\acmConference[ICSE '26]{ICSE '26}{April 12--18,
  2026}{Rio de Janeiro, Brazil}
\acmISBN{978-1-4503-XXXX-X/2018/06}

\usepackage{amsmath}
\usepackage{multirow}
\usepackage{float}
\usepackage{makecell}
\usepackage{dirtytalk}
\usepackage{graphicx}
\usepackage{xcolor}
\newenvironment{sayit}[1]
{\textit{\say{#1}}}

\copyrightyear{2026}
\acmYear{2026}
\setcopyright{cc}
\setcctype{by}
\acmConference[ICSE '26]{2026 IEEE/ACM 48th International Conference on Software Engineering}{April 12--18, 2026}{Rio de Janeiro, Brazil}
\acmBooktitle{2026 IEEE/ACM 48th International Conference on Software Engineering (ICSE '26), April 12--18, 2026, Rio de Janeiro, Brazil}
\acmPrice{}
\acmDOI{10.1145/3744916.3773172}
\acmISBN{979-8-4007-2025-3/2026/04}

\begin{document}

\title[GenAI Use by Blind and Low Vision Software Professionals in the Workplace]{``Game Changer" or "Overenthusiastic Drunk Acquaintance"? Generative AI Use by Blind and Low Vision Software Professionals in the Workplace}

\author{Yoonha Cha}
\email{yoonha.cha@uci.edu}
\affiliation{%
  \institution{University of California, Irvine}
  \city{Irvine}
  \state{CA}
  \country{USA}
  }

\author{Victoria Jackson}
\email{v.jackson@soton.ac.uk}
\affiliation{%
  \institution{University of Southampton}
  \city{Southampton}
  \state{}
  \country{UK}
}

\author{Lauren Shu}
\email{lpshu@uci.edu}
\affiliation{%
  \institution{University of California, Irvine}
  \city{Irvine}
  \state{CA}
  \country{USA}
  }

\author{Stacy M. Branham}
\email{sbranham@uci.edu}
\affiliation{%
  \institution{University of California, Irvine}
  \city{Irvine}
  \state{CA}
  \country{USA}
  }

\author{André van der Hoek}
\email{andre@ics.uci.edu}
\affiliation{%
  \institution{University of California, Irvine}
  \city{Irvine}
  \state{CA}
  \country{USA}
  }

\renewcommand{\shortauthors}{Yoonha Cha et al.}

\begin{abstract}
The software development workplace poses numerous technical and collaborative accessibility challenges for blind and low vision software professionals (BLVSPs). Though Generative AI (GenAI) is increasingly adopted within the software development industry and has been a rapidly growing topic of interest in research, to date, the unique perspectives of BLVSPs have yet to be consulted. We report on a qualitative study involving 39 semi-structured interviews with BLVSPs about what the introduction of GenAI has meant for their work. We found that BLVSPs used GenAI for many software development tasks, resulting in benefits such as increased productivity and accessibility. However, significant costs were also accompanied by GenAI use as they were more vulnerable to hallucinations than their sighted colleagues. Sometimes, organizational policies prevented use. Based on our findings, we discuss the higher-risks and higher-returns that BLVSPs had to carefully weigh when deciding whether and when to use GenAI tools for work.

\end{abstract}



\maketitle

\section{Introduction}

According to the most recent Stack Overflow survey that includes disability demographics (2022)~\cite{stackoverflow_2022}, 1.7\% percent of the total respondents reported being blind or having low vision (BLV), comprising 44.8\% of the respondents who identified as having a disability. Existing scholarship regarding blind and low vision software professionals (BLVSPs) documents the significant labor they invest into gaining access at work, as many tools and processes used for daily job tasks (e.g., IDEs, project management tools) remain inaccessible~\cite{das_ideation_24, pandey_programming_21, alharbi_accessibility_2023, filho_visual_2015, albusay_code_navigation_17}. This additional, \say{invisible work}~\cite{branham_invisible_2015} materializes in many different forms, including procuring assistance from sighted coworkers~\cite{albusay_code_navigation_17}, identifying alternative accessible alternative solutions~\cite{huff_workexp_20, cha_participate_2024, pandey_understanding_2021}, and even building do-it-yourself (DIY) tools outside of regular work hours~\cite{cha_dilemma_2025}. As a result, BLVSPs experience inequitable barriers to lateral and hierarchical career progression~\cite{cha_understanding_2024}. 


In the software development profession, Generative Artificial Intelligence (GenAI) has gained significant attention and is rapidly becoming integrated into professionals' everyday workflows~\cite{alphabet25}. 76\% of all respondents to Stack Overflow's 2024 survey indicated an intention to use GenAI tools for software development, and 62\% reported already utilizing them; this marks a 6\% and 18\% increase, respectively, from the previous year~\cite{stackoverflow_2024}. Much research has focused on the use of GenAI in programming~\cite{hou_large_2024}, maintenance~\cite{hou_large_2024}, and testing~\cite{hou_large_2024}, how to efficiently integrate GenAI tools into software development~\cite{russo_navigating_2024}, and its impact on the job market~\cite{malheiros_impact_2024, lima_generative_2024}. However, while GenAI is being taken up by software professionals to support productivity and efficiency
, little is known about the experiences of underrepresented groups in software development with GenAI tools and how they may benefit or hinder them. In particular, the experiences of BLVSPs with GenAI and the potential of this technology to reduce the burden of invisible work and overcome career barriers have yet to be explored. To address this gap, we investigated the following research questions:
\begin{itemize}
    \item RQ1. How do BLVSPs perceive and utilize generative AI in the workplace?
    \item RQ2. How do generative AI tools impact BLVSPs' work and career prospects?
\end{itemize}


We report the results from in-depth, semi-structured interviews with 39 BLVSPs who identified as being blind or having low vision, are working or have worked in a software development position either in a corporate setting or as a freelancer, and have at least one year of work experience. Through thematic analysis~\cite{braun_using_2006}, we identified three main themes: (1) benefits of using GenAI, (2) barriers and risks associated with using GenAI for work, and (3) the wider impact that GenAI has on BLVSPs in the workplace. We discuss BLVSPs' rationales for deciding whether and when to use GenAI tools at work, and suggest implications for GenAI tool developers, researchers, and employers.

Through this study, we contribute the following:
\begin{itemize}
    \item A comprehensive understanding of BLVSPs' decisions about the (non) use of Generative AI tools in the workplace.
    \item Impact of GenAI tools on overall workplace accessibility and career paths for BLVSPs.
    \item A set of implications for GenAI tool developers, researchers, and employers.
\end{itemize}

\section{\textbf{Background and }Related Work}

\subsection{Assistive Technology}
As per the World Health Organization, \say{Assistive Technology (AT) enables and promotes inclusion and participation, especially of persons with disability, aging populations, and people with non-communicable diseases}~\cite{WHO}. Popular AT products used by BLV individuals include screen readers (e.g., NVDA~\cite{nvda}, JAWS~\cite{jaws}) that speak the content of a webpage or application out loud and are navigable with key combinations, screen magnifiers that zoom in and out to help read the contents of a screen, and braille displays that enable access to information by raising and lowering pins in refreshable, tactile braille character cells.

\subsection{BLVSPs in the Workplace}
BLVSPs face challenges in equitable participation in the workplace due to inaccessible developer tools~\cite{pandey_understanding_2021}, work environments~\cite{cha_understanding_2024, cha_participate_2024}, and ableist colleagues~\cite{das_win_friends_19, cha_understanding_2024}. Overcoming these barriers requires additional access labor of negotiating practices, and developing workarounds and bespoke tools to be able to complete job tasks~\cite{cha_participate_2024, cha_understanding_2024, cha_dilemma_2025}. Legislation and policies in many countries require companies to provide reasonable accommodations (e.g., the Americans with Disabilities Act (ADA) in the USA~\cite{ADA.gov}). Such accommodations include the provision of adequate ATs and flexible work practices. 
Yet, accommodations are not always made in a timely manner, leading to workarounds and frustrations~\cite{marathe_accessibility_2025}. The workplace continues to remain challenging for many BLVSPs~\cite{cha_understanding_2024}, leading to career advancement concerns~\cite{cha_understanding_2024, das_ideation_24}.  

While software development is a collaborative endeavor~\cite{mistrik2010collaborative}, it can be challenging for BLVSPs to fully participate in collaborative work (e.g.,~\cite{huff_workexp_20, pandey_programming_21}). Remote meetings introduce difficulties such as the need to juggle the use of a screen reader to interpret a presentation while simultaneously listening to the presenter~\cite{tang_telework_21}. Ideating with colleagues is tricky, as popular digital whiteboarding tools (e.g., Miro) do not work well with screen readers or braille displays~\cite{das_ideation_24}. Collaborative authoring imposes additional work as collaborative document editors, while accessible, are often not usable for BLVSPs~\cite{das_win_friends_19}. BLVSPs thus rely on workarounds such as assistance from sighted colleagues (e.g.,~\cite{das_win_friends_19, tang_telework_21, das_ideation_24}) or custom tools~\cite{cha_dilemma_2025} to collaborate effectively. Deciding whether to ask colleagues for assistance is often stressful for BLVPs as asking for help discloses their disability with worries about how this will be perceived and potential negative consequences~\cite{cha_participate_2024}.

A key area of research focus has been the accessibility of technical tasks such as authoring code (e.g., ~\cite{albusay_code_navigation_17, armaly_audiohighlight_18, pandey_programming_21, pandey_towards_2024, potluri_codetalk_18}), pair-programming (e.g., ~\cite{pandey_programming_21}), and debugging (e.g.~\cite{potluri_codetalk_18, stefik_sodbeans_2009}). Among the challenges caused by accessibility issues are difficulties in navigating code~\cite{albusay_code_navigation_17}, reading code in an IDE~\cite{pandey_towards_2024}, and limited pair programming contributions~\cite{pandey_programming_21}. To improve the accessibility of technical tasks, various IDE plugins have been proposed by researchers (e.g.,~\cite{potluri_codewalk_2022, armaly_audiohighlight_18}. For example, CodeTalk~\cite{potluri_codetalk_18} summarizes information within an IDE screen as an audio feed. 
More recently, tools that incorporate AI to improve accessibility, such as a command-line debugging tool~\cite{saben_enabling_2024}, have been explored.

To support the completion of technical tasks, developers (including BLVSPs) seek information from forums, blogs, tutorials, and official documentation websites. 
Many of these sources pose accessibility issues that require workarounds, such as custom scripts or asking sighted colleagues for assistance~\cite{storer_Infoseeking_21, cha_dilemma_2025}.

\subsection{GenAI Use in Software Development}
Developers have rapidly integrated GenAI \cite{stackoverflow_2024} into numerous development tasks, including code generation~\cite{pereira2025exploring}, authoring test cases~\cite{sergeyuk2025using}, and debugging~\cite{khojah_beyond_2024}. One of the primary benefits of adopting GenAI is increased developer productivity (e.g., \cite{bird_copilot, ziegler2024measuring, pereira2025exploring}). This increase is due to reasons such as GenAI enabling users to quickly search for pertinent information~\cite {davilaIndustryCaseStudy2024}, automating boilerplate code~\cite{kemell_still_2025}, and auto-completing code~\cite{mendes_ai_code_assistant_2024}. Use of GenAI has also been noted to increase job satisfaction~\cite{sergeyuk2025using}. However, GenAI can generate poor quality code~\cite{liu_refining_code_quality}, insecure code~\cite{perry_insecure_code_2023}, and inaccessible code~\cite{aljedaani_acc_chatgpt_2024}, for which rework is needed to fix the mistakes~\cite{liangLargeScaleSurveyUsability2024}. Trust in GenAI tools is a significant concern among developers (e.g.,~\cite{khojah_beyond_2024, wangInvestigatingDesigningTrust2023}) as well as skepticism about how well GenAI can assist with more complex software development tasks~\cite{khemka_AI_fordevs}.

GenAI has shown potential to improve software accessibility for BLV people by assisting developers in authoring accessible applications (e.g., ~\cite{aljedaani_acc_chatgpt_2024, cali_prototype_2025, mowar_codeal11y_2025, suh_human_llm_acc_2025, yu_llm_screen_reader_2025}). 
One study found that GenAI-generated code has fewer accessibility errors in some accessibility categories than human-generated code, but in other categories it introduces more accessibility errors than humans~\cite{suh_human_llm_acc_2025}. Another noted that GenAI was able to fix some, but not all, accessibility errors in both GenAI and human-generated code~\cite{aljedaani_acc_chatgpt_2024}. To better support developers, researchers have explored methods to facilitate the creation of accessible code. For example, MS VS Code plugins use GenAI to automatically detect accessibility issues in the code ~\cite{mowar_codeal11y_2025}, alert developers of accessibility issues~\cite{mowar_codeal11y_2025}, and suggest fixes to identified issues~\cite{cali_prototype_2025}. While the plugins result in fewer accessibility errors, more complex issues still require developer expertise to address~\cite{mowar_codeal11y_2025}. While GenAI shows promise, developer education and training remain vital for developing accessible applications~\cite{mowar_codeal11y_2025}.

As GenAI use becomes commonplace and an increasing amount of coding is automated~\cite{alphabet25}, it is unclear what impact GenAI will have on the developer profession~\cite{economist_swdev_2024} and the nature of development work. Some project that the demand for AI-based products will increase job opportunities for developers~\cite{USBLS_2025, WEF_Future_Jobs_2025}, so it is critical for developers to upskill by learning how to use GenAI effectively~\cite{kam_professional_2025}. Early indications are that development work is becoming more rote with the potential to reduce job satisfaction~\cite{amazon_nyt2025}. Although others believe developers will need to focus on the more complex tasks AI struggles with~\cite{economist_swdev_2024, kang2024}, or the more interesting, valuable creative tasks~\cite{jackson_genai_creativity}. Finally, the impact of GenAI adoption on the software profession may not be equally distributed, since developers in minority groups may be more negatively impacted~\cite{hicks_new_2024}.

\subsection{GenAI Use by BLV People}
GenAI tools are used by BLV people to describe visual information contained within images or photos~\cite{adnin_blind_genai_2024, tang_everyday_uncertainty}. The rich descriptions provided are often superior to those offered by sighted individuals. However, hallucinations cause BLV people to contemplate how much they can trust the response~\cite{tang_everyday_uncertainty} and whether verification through other sources (e.g., using Google, prompting multiple models, asking another person) is required~\cite{tang_everyday_uncertainty}. This decision to verify depends on the context of use, how high the stakes are, whether it can be verified, and relying on perceived ``believability'' of the response~\cite{adnin_blind_genai_2024}, which requires additional work from BLV people. Furthermore, to even use GenAI tools, individuals often need to overcome accessibility issues present in GenAI tools. For example, buttons are unlabeled, or headings and landmarks are not present in ChatGPT, making it difficult to use the tool with a screen reader~\cite{adnin_blind_genai_2024}. This inaccessibility in GenAI applications leads to concerns that people with disabilities (PWD)---including BLV people---will be left behind and unable to reap the many benefits promised by GenAI~\cite{alshaigy_forgotten_2024}. 

Beyond the multi-purpose GenAI tools used for both general work and development work, such as ChatGPT and Claude, developers also utilize specific developer tools like GitHub Copilot, which leverages AI to suggest code as they author it. A recent study~\cite{flores_impact_2025} on the accessibility of GitHub Copilot found that BLV programmers experienced similar benefits as sighted developers, including improved productivity. However, Copilot also posed challenges and frustrations to BLVSPs, such as constant context switching between tabs, breaking the coding flow, and requiring multiple keystrokes to navigate through the application. 



This study contributes to existing scholarship on the experiences of BLVSPs at work by investigating their experiences with using various GenAI tools beyond GitHub Copilot to support better workplace accessibility.


\section{Study Design} \label{study_design}

\begin{table*}[ht!]
\centering
\begin{tabular}{|p{0.2\linewidth}|p{0.5\linewidth}|p{0.1\linewidth}|p{0.1\linewidth}|}
\hline
\textbf{Higher-Level Themes} & \textbf{Themes} & \textbf{RQs} & \textbf{Section} \\ \hline
\multirow{6}{*}{\parbox{1.3in}{Benefits of GenAI}} & Productivity Gains and Well-Being & \multirow{6}{*}{RQ1, RQ2} & \multirow{5}{*}{4.2}\\ \cline{2-2}
 & Less Reliance on Sighted Colleagues &  &\\ \cline{2-2}
 & Reduced Access Labor - Creation of or Dispensation with DIY Tools &  &\\ \cline{2-2}
 & Better Participation in Professional Conversations &  &\\ \cline{2-2}
 & Maintaining Professional Images &  &\\ \hline
\multirow{4}{*}{Costs and Risks of GenAI} & Suboptimal UX of GenAI Tools & \multirow{4}{*}{RQ1, RQ2} & \multirow{4}{*}{4.3} \\ \cline{2-2}
 & Meticulous Validations & & \\ \cline{2-2}
 & Increased Vulnerability & & \\ \cline{2-2}
 & Non-Use vs. Proceeding Despite the Risks & & \\ \hline
\multirow{5}{*}{\parbox{1.3in}{Wider Impact of GenAI on Workplace Equity}} & Benefiting or Harming Accessibility (Awareness) in the Workplace & \multirow{5}{*}{RQ2} & \multirow{5}{*}{4.4}\\ \cline{2-2}
 & Creating More Positions for BLVSPs & & \\ \cline{2-2}
 & Supporting BLV People's Pursuit of Software Development & & \\ \cline{2-2}
 & Being Recognized as a Competitive Candidate for Promotions & & \\ \cline{2-2}
 & Organizational Policies around GenAI Use & & \\ \hline
\end{tabular}%
\centering
\caption{The higher-level themes and themes along with corresponding research questions and findings section.}
\label{tab:themes_rqs}
\end{table*}

This study builds on an existing study examining how BLVSPs 
overcome accessibility issues at work by building and using do-it-yourself (DIY) tools~\cite{cha_dilemma_2025}, in which we asked several initial questions about their general engagement with GenAI. This provided us with some early insights, though these did not cover all aspects of interest for the researchers. Thus, the researchers decided to conduct a further set of interviews to probe more specifically into BLVSPs' GenAI use at work.


\subsection{Semi-Structured Interviews}
We prepared a new interview protocol, where we kept some GenAI-related questions from the previous protocol, and  added new questions of interest. These questions were informed by prior studies on the adoption of GenAI by developers (e.g., ~\cite{davilaIndustryCaseStudy2024, pereira2025exploring}) and the researchers' prior accessibility-related studies. The original protocol (questions related to GenAI are highlighted) and the new interview protocol are available in supplementary materials~\cite{Dryad_SuppData}.

\subsection{Participants} \label{sec:demographics}
We sought to recruit software professionals who: (1) self-identified as being blind or having low vision, (2) working or have worked as a software professional (e.g., software engineer, accessibility specialist, product manager, designer), and (3) have at least one year of professional work experience. 
We employed several non-probabilistic approaches~\cite{baltes2022sampling} to recruitment. Firstly, we used the researchers' professional networks to connect with existing contacts and potential participants, resulting in 14 recruits. We also sent recruitment emails to mailing lists tailored to BLVSPs such as Program-L~\cite{program-l}, which resulted in the recruitment of 16 participants. Finally, 9 participants were recruited via snowball sampling~\cite{baltes2022sampling}.

Our participants had diverse demographics. First, they reported a spectrum of vision abilities. 20 self-reported as being totally or completely blind (51\%), 6 as having light and / or color perceptions (15\%), and 13 participants (33\%) as having varying degrees of residual vision, such as tunnel vision and 2\% useful vision. Participants were located in various countries across continents, with 67\% from North America (n = 26) and the rest being from Europe (n = 6), Asia (n = 4), and South America (n = 3). Participants had a wide range of work experiences; 17 had 1 to 5 years, 10 had 6 to 10 years, 5 had 11 to 20 years, 3 had 21 to 30 years, and 4 had 31 years or more years of experience working. Company sizes also varied, from startups to multinational corporations. 34 participants identified as men, 4 as women, and 1 as non-binary. Participants had various job positions, the most common being software engineers (n = 21), followed by accessibility specialists (n = 5) and other positions such as UX specialists and technical executives. 
Participant IDs were randomly assigned, and we selectively omit participant IDs in places where disclosure may lead to potential participant identification. Further participant information is specified in supplementary materials~\cite{Dryad_SuppData}.


\subsection{Procedure}

We conducted audio-recorded, semi-structured interviews with 39 BLVSPs over Zoom. 
30 participants were interviewed as part of the prior study on DIY tools~\cite{cha_dilemma_2025}, between May 2024 and July 2024. 9 additional participants were recruited specifically for this study on the use of GenAI at work, interviewed between January 2025 and February 2025. During this period, we also re-interviewed 5 of the 30 prior participants who responded to a request for additional interviews. While there was a difference of approximately six months between the times of the first and second set of interviews, the user interfaces (UI) of the tools used did not change radically and hallucination was a well-known concept throughout. This paper presents the results from the collective interviews with 39 BLVSPs. 
37 interviews were conducted in English. Two interviews were conducted in Portuguese as preferred by the participants, by researchers whose native language was Portuguese.

Interviews ranged from 43 minutes to 182 minutes, with an average of 72 minutes. Participants were emailed a study information sheet for review and completed a form collecting demographic information. Researchers acquired verbal consent at the beginning of each interview. Questions asked during the interviews covered multiple topics, including the GenAI tools used, the tasks for which they are used, and their opinions on using GenAI tools for work. Interview protocols are available in supplementary materials~\cite{Dryad_SuppData}. Participants were compensated at a rate of \$40 per hour via an Amazon gift card or a similar method.

\subsection{Ethics and Data Collection}
This study was approved by the researchers’ Institutional Review Board (IRB). 
All audio recordings and transcripts were stored on a secure drive accessible only to the researchers. All personally identifiable information, such as participant names, was removed from the transcripts before data analysis.

\subsection{Data Analysis}
All recorded audio files were transcribed and verified for accuracy by the research team. The two audio recordings in Portuguese were transcribed and translated into English by the researcher undertaking the interview, who is a native Portuguese and a proficient English speaker. We first performed thematic analysis~\cite{braun_using_2006} to analyze the data. The first two authors analyzed the same five randomly selected transcripts, performed open coding, and frequently met to discuss and identify similarities and differences between themes. After discussions with the identified and refined themes and reaching an agreement, the first author revisited the five transcripts to analyze them for new themes. Then, the first author coded the remaining 34 transcripts to identify themes. The identified themes were presented to the entire research team in regular meetings and further discussed. Higher-level themes were identified by the team through this process. These themes are presented in~\autoref{tab:themes_rqs}. 


We did not adopt coding reliability measures such as inter-rater reliability (IRR). IRR is a statistical measure aiming to ensure ``objective'' and ``unbiased'' coding, rooted in positivism~\cite{braun_one_2021, mcdonald_reliability_2019}. However, in reflexive and interpretivist approaches to qualitative research, which we ascribe to, introducing coding reliability is ``illogical''~\cite{braun_one_2021}, as reflexive thematic analysis considers researcher subjectivity and interpretation as valuable analytical resources~\cite{braun_one_2021}. Therefore, we did not calculate IRR. This is also seen in other research in software engineering~\cite{pimenova_good_2025, culas_newcomers_2025}.

Additionally, two researchers independently reviewed all transcripts to identify the various GenAI tools used and the tasks they supported (see~\autoref{41}). In multiple half-hour meetings, two researchers reviewed and discussed results met with a third researcher who has extensive experience in the software engineering industry, to discuss alignments and divergence, and collectively make adjustments.



We did not seek complete saturation~\cite{saunders_saturation_2018} in conducting interviews. Recruiting more participants for saturation was challenging, as (1) the population size of BLVSPs is relatively small compared to sighted software professionals, (2) PWD and marginalized groups are already overburdened, with the small population already being constantly asked by researchers to participate in studies~\cite{dee_pool_2016}, and (3) this population is already shouldering significant access labor~\cite{branham_invisible_2015, cha_participate_2024, cha_understanding_2024} to make their work accessible. 





\section{Findings}
In each subsection, each high-level theme identified by the researchers is addressed. \autoref{tab:themes_rqs} lists the research questions each higher-level theme answers. 
Section \ref{41} describes the use cases, and sections \ref{42} and \ref{43} describe the benefits and drawbacks of GenAI use. We note that in some use cases, we observed both benefits and drawbacks. Lastly, \autoref{44} documents the impact of GenAI tools on BLVSPs' work experiences on a higher level. 

\begin{table*}[t]
\centering
\resizebox{\textwidth}{!}{%
\begin{tabular}
{|p{1.7in}|p{4.8in}|}
\hline
\textbf{Tasks} & \textbf{Examples} \\ \hline
\multirow{2}{*}{Ideation / Brainstorming} & Image, diagram, slide generation \\ \cline{2-2} 
 & Rubber-ducking solutions \\ \hline
\multirow{2}{*}{Architecture / Design} & Identifying suitable libraries \\ \cline{2-2} 
 & Making architectural decisions \\ \hline
\multirow{3}{*}{Producing Code} & Writing unit test cases \\ \cline{2-2} 
& Writing code in familiar languages (e.g., Python, C++)\\ \cline{2-2}
 & Writing code in an unfamiliar language (e.g., YAML, PHP) \\ \hline
\multirow{6}{*}{Understanding Code} & Learning an unfamiliar programming language \\ \cline{2-2} 
 & Navigating inaccessible Maven files in the IDE \\ \cline{2-2} 
 & Code review (e.g., Reviewing pull requests and ``diff'' files) \\ \cline{2-2} 
 & Skimming lengthy legacy / unfamiliar codebases (e.g., onboarding) \\ \cline{2-2} 
 & Debugging (e.g., log file analysis, pinpointing errors on certain lines) \\ \cline{2-2} 
 & Asking theoretical algorithmic questions (e.g., Dynamic Programming or Greedy Solution) \\ \hline
\multirow{4}{*}{Visual Descriptions} & Describing screenshots of code or product UI sent by colleagues \\ \cline{2-2} 
 & Describing technical visual materials (e.g., UML diagrams, flow charts) \\ \cline{2-2} 
 & Describing inaccessibly formatted tables, slides, and images \\ \cline{2-2} 
 & Verifying rendered GUI during front-end development \\ \hline
\multirow{2}{*}{\parbox{1.2in}{Creating and Delivering Presentations}} & Checking that presentation is visually accessible to sighted colleagues \\ \cline{2-2} 
 & Facing directly into the camera during video conferencing or video recording \\ \hline
\multirow{5}{*}{General Information Seeking} & Recalling best practices, naming practices, software principles, etc. \\ \cline{2-2} 
 & Summarizing information that would otherwise require multiple (inaccessible) web searches \\ \cline{2-2} 
 & Skimming printed material documents to assess need for full access \\ \cline{2-2} 
 & Summarizing WCAG guidelines for colleagues \\ \cline{2-2} 
 & Querying shortcuts for software development tools such as GitHub and GitLab \\ \hline
\multirow{3}{*}{\parbox{1.7in}{Locating Information within Codebases and Documentations}} & Generating python code that summarizes educational materials without traversing each page 
\\ \cline{2-2} 
 & Detecting trends and anomalies in the data file 
 \\ \cline{2-2} 
 & Data analysis - finding a specific key-value pair in a large JSON file \\ \hline
Miscellaneous & Troubleshooting when AT shuts down \\ \hline
\end{tabular}%
}
\caption{Tasks BLVSPs used GenAI tools for and example cases.}
\label{tab:usecases}
\end{table*}
\subsection{GenAI Usage by BLVSPs} \label{41}
Just like their sighted counterparts, BLVSPs were enthusiastic about GenAI, which \sayit{ma[de] [their] life much easier} (P33), enabling them to \textit{``do things [tasks] more efficiently, and not flail around''} (P16). Most participants praised these tools as a \textit{``super competitive''} (P35) workaround, bringing them closer to equitable workplace experiences. Participants were fond of GenAI: \textit{``I feel it's probably the most reliable, consistent and affordable solution compared to whatever I've used before''} (P35).


Participants used a range of GenAI tools for work, with varying frequency of use. All participants answered having used GenAI in some capacity. 62.5\% (n = 25) used it daily, sometimes multiple times a day, while the remaining 37.5\% (n = 15) only occasionally used it when deemed necessary for work or out of \textit{``general curiosity''} (P11). Frequently used GenAI tools included Claude (3.5 Sonnet, 3.5 Haiku)~\cite{Claude}, DeepSeek~\cite{DeepSeek}, and Midjourney~\cite{Midjourney}, with ChatGPT (GPT-4, GPT-4o)~\cite{chatgpt}, Gemini~\cite{gemini}, and Copilot~\cite{copilot} being the most used. 
Notably, BLVSPs actively used GenAI tools \textit{``specifically designed for... visually impaired people... [that] help me work with [existing software]'}' (P37) such as ``Be My AI'' and ``Picture Smart'' (JAWS feature). Five participants even created their own GenAI-based DIY tools to meet specific needs.

GenAI assisted BLVSPs with numerous software development tasks. Like their sighted co-workers, they used it to write code and boost their productivity in doing so. In addition, GenAI tools helped with tasks that are often \textit{``pretty visual''} (P7), including programming tasks like debugging and assisting with ingesting or creating visual materials like flow charts, tables, diagrams, and presentations. P1 described a major shift in his approach for handling inaccessible visuals: \textit{``Recently I dropped everything else and now I'm using LLMs to take pictures and get image descriptions.''} \autoref{tab:usecases} is a comprehensive list of tasks our participants reported to use GenAI for. Due to limited space, we only elaborate on a few select tasks below.

\subsubsection{\textbf{Debugging}}
GenAI tools greatly improved the accessibility of debugging, which traditionally poses accessibility challenges due to highly-visual error indicators and the complex structure of terminal output~\cite{potluri_codetalk_18}. For instance, P38 described using ChatGPT to interpret the \textit{``confusing''} output of the Linux command ``psdash'': \textit{``It [psdash]'s going to show you the output in tabular format. So it looks like a table... but a screen reader cannot detect which is the process ID or which is the running time and so on... it's hard to figure [that] out.''} Before the introduction of GenAI tools, he \textit{``had to remember each column name, number of columns, and then the order, and then try to read it word by word.''} To tackle this inconvenience, he decided to simply \textit{``copy the terminal output, paste it into the GPT, and then... ask a specific question that I have.''} Similarly, BLVSPs used ChatGPT to avoid traversing lines of code with a screen reader to locate an error or \textit{``write a program to find it[error]''} (P13), when \textit{``a sighted person can just look at the screen and find it''} (P13): 
\begin{quote}
    \sayit{Sometimes it's 1000 lines. [I have to go through] character by character... It would take a whole day just to check that code. Then I paste it in ChatGPT and it says, `There are missing parentheses on line 998.' I fix it and it works.} (P21)
\end{quote}

\subsubsection{\textbf{Front-End Development Verification}}
GenAI tools empowered BLVSPs to perform visual development tasks: particularly, front-end UI development and review. For example, P13 described utilizing GenAI to validate whether his output was \textit{``visually attractive.''} These tools were also used to enable participation in UI design review sessions, which rely on highly visual tools such as Figma, \textit{``a staple in every company in the SDLC''} (P7):
\begin{quote}
   \textit{``Before [Gen]AI, my option in the design review was to say, `You're going to need to describe to me everything that's there, and everything that's not there.' Which is, a huge gotcha. You can't explain to me something that's not there if you don't know what's supposed to be there. So, I can use [Gen]AI to criticize designs that people have, and ask if things exist.'' (P7)}
\end{quote}

\subsubsection{\textbf{Interpreting Documentation}}
GenAI also helped our participants understand inaccessible materials, such as API documentation in improperly tagged PDFs. Several participants chose to use GenAI to ask for a description of the PDF (P35) or to restructure it into more accessible HTML or Markdown (P13). 

\subsubsection{\textbf{Interpreting Technical Visuals}}
Almost all of our participants used GenAI to help interpret flow charts, diagrams, tables, and even screenshots of code, acknowledging its convenience and speed: \textit{``You just click two buttons; open the targeting menu, choose the model, and it gives you a description... A two-key combination is pretty fast and pretty convenient"} (P15). For straightforward visuals such as tables and bar charts, GenAI proved highly effective, with our participants' confidence level in using GenAI for these being \textit{"80\% or 90\%, something like that"} (P13). However, BLVSPs mostly found GenAI less helpful for relatively complex visuals like flow charts, especially when they did not have enough context beforehand: \textit{``Flowchart is the worst one, like a process chart... If I have no [prior] idea what that is, then it's very hard to understand''} (P33). 

GenAI was also \textit{``good for recognizing screenshots that [sighted colleagues] send me''} (P15), usually of product UI and code snippets. P32 utilized ChatGPT to interpret the screenshots of code without alternative text and \textit{``press my magical key combination''} to retrieve the code on the image. 
Another perk of GenAI tools that other workarounds did not have, was being able to \textit{``dig deeper''} (P34) about responses when asking for such visual descriptions, without taking up too much of their colleagues' time or feeling pressured to ask all the questions within a limited time frame.


\subsection{Benefits of Using GenAI} \label{42}
By using GenAI tools in many aspects of their work, participants' reliance on sighted colleagues greatly decreased \textbf{when compared with prior workflows}, while maintaining a professional image and experiencing a positive impact on their mental well-being.

\subsubsection{\textbf{Productivity Gains and Well-Being}}
Our participants believed that \textit{``GenAI makes blind engineers much more productive''} (P34), noting additional productivity boosts and enhanced mental well-being. Traditionally, inaccessible tools \textit{``impacted [their] performance negatively''} (P12) and \textit{``definitely takes a mental toll on you... a really bad one''} (P12). BLVSPs reported substantial improvements in their work efficiency and stress levels after using GenAI: \textit{``I'm faster and I use less resources. I'm not as stressed''} (P10). For many, the benefits justified paying for premium subscriptions because \textit{``it just takes me so much longer to do my job without [Gen]AI''} (P14). 

\subsubsection{\textbf{Less Reliance on Sighted Colleagues}}
Despite needing sighted assistance at times, participants felt uneasy burdening their colleagues. Sometimes, this discomfort stemmed from the desire to maintain independence and the fear that their accessibility requests may be misconstrued as excessive reliance. They were also conscious of the cumulative impact of repeated requests on their sighted colleagues' productivity: 
\begin{quote}
    \textit{``If you're having to repeatedly, constantly ask sighted colleagues, you can take the amount of time that you are both spending to solve this problem, you can add your rate and the person's rate that you're asking the question to... And the more times that that happens, the more difficult it becomes to ask for help.''} (P7)
\end{quote}

Thus, participants perceived GenAI as \textit{``the colleague that doesn't get annoyed at me''} (P7) and used GenAI tools in doing their due diligence before seeking human intervention---getting a general grasp of the task, doing their best to work around most accessibility issues---and only going to colleagues for support when absolutely necessary so the colleagues are \textit{``more inclined to help''} (P13). Even when assistance from \textit{``human beings can do a much better job''} (P13), participants described actively trying to \textit{``minimize it''} (P13). 

\subsubsection{\textbf{Reduced Access Labor - Creation of or Replacement for DIY Tools}}
GenAI streamlined the DIY process and in some cases removed the need to DIY altogether. This substantially lightened the burden on our participants who previously invested considerable personal time and energy---often outside of regular work hours---crafting custom tools to navigate workplace accessibility challenges. 

Many BLVSPs recognized the benefits of DIY-ing accessible solutions in their careers, but also emphasized the exhaustion the process caused, as \textit{``it is a commitment to build the tools''} (P16). They also preferred to keep work and personal lives separate and invest in personal growth and passion projects outside of work. Also commonly shared among participants was the fatigue coming from extra work: \textit{``I'm very busy and tired... I'm just trying to get my baby fed and find litter for my cat"} (P16).

GenAI tools proved particularly useful in assisting BLVSPs with the ideation, brainstorming, and coding process of the DIY tool. In addition, when participants were not familiar with a particular language to be coding the DIY solution from scratch, oftentimes GenAI tools supported the swift authoring of the required solution. P10 shared an instance of utilizing ChatGPT to write a JavaScript code snippet---a language he \textit{``really did suck at''}---to improve usability of Skype in supporting seamless alternating between the chat and the main call was troublesome. 


While GenAI could not resolve every use case for the need to DIY, participants largely shared the notion that GenAI proved a useful addition to their toolbox that decreased the \textit{``unwanted access labor''} (P38) often required in their work: \textit{``Before GenAI it was just the two things, right? Ask a sighted person or build your own tool... It is still semi-frequent. But it's significantly less than before''} (P13).

\subsubsection{\textbf{Affordable Alternative to Unfeasible AT}}
The paid subscription plans also turned out to be a much more affordable alternative to accessible tools like Braille Displays that were \textit{``extremely unfeasible options''} (P21), which P21 highlighted can cost \textit{``around R\$ 20,000 in Brazil (around USD \$4,000)... It would make my life easier... [But] I'm not going to spend R\$20,000''} (P21).

\subsubsection{\textbf{Better Participation in Professional Interactions}}
Participants acknowledged that descriptions of work materials, such as diagrams and flowcharts obtained through GenAI tools, led to more engaging and comprehensive professional interactions with colleagues compared to their interactions prior to the introduction of GenAI tools. P24 shared: \textit{``I get to have conversations that I've never had with people before... that feels quite liberating''} (P24). Additionally, P35 was able to \textit{``talk a little bit more visually''} at work with information gained from GenAI tools about UIs and screen layouts:
\begin{quote}
    \textit{``I've never seen a computer screen... Before, I would be super linear as to how a screen reader would [describe it] to me... And now I'm able to speak to it a little more confidently, in terminology that they [sighted colleagues] are more familiar with.''} (P35)
\end{quote}

\subsubsection{\textbf{Maintaining Professional Image}}

As GenAI increased productivity and well being, and reduced sighted assistance and excessive access labor and provided affordability, 
BLVSPs were able to better manage their professional image and participate better in professional interactions. P24 shared, \textit{``on paper, it looks like I'm a lot better because the tester just doesn't get any incomplete pieces of work from me anymore.''}  
While BLVSPs used to work overtime in concern that colleagues may be `\textit{``inclined to think that I'm not giving this project the time that it deserves''} (P13) if they did not, P13 felt that \textit{``AI has definitely helped''} maintain his professional image with less access labor. 

Notably, GenAI provided protection against unwanted disability disclosure in professional settings. P7 recounted having clients who \textit{``don't even know that I'm blind... I prefer to keep it that way because I don't think it detracts from the quality of my work.''} Further, P13 shared his strong preference for not disclosing his disability due to numerous prior experiences where he had \textit{``lost projects just by sharing the fact [that I'm blind]. Many, many, many times... someone said, ‘Since you're visually impaired, you won't be able to [work],' and they just terminated my contract in a few days.''} Both participants praised GenAI's safeguarding from inadvertent disclosure: \textit{``Now it's easier for that [my blindness] not to even be brought up in the first place because if there's something that's visual in nature, OK, let me just have AI do it for me''} (P7). P13 took a step further in being extra careful to not even inadvertently give away hints at his disability by using AI to reformat code in a way that other sighted developers would do.


\subsection{Barriers and Risks of Using GenAI} \label{43}
While we saw benefits of using GenAI tools, BLVSPs also took significant risks. 
In addition to general concerns like hallucinations---which took longer for BLVSPs to detect and correct---additional costs such as inaccessibility and low usability further eroded GenAI's potential advantages. These risks and costs compounded, and often led BLVSPs to either abandon GenAI tools for already established workarounds and DIY solutions or meticulously weigh benefits against these considerable costs.

\subsubsection{\textbf{Suboptimal UX of GenAI Tools}}
BLVSPs were stressed about the dissatisfactory UX and accessibility of GenAI tools, reported feeling left behind, and shared coping strategies. 

\paragraph{Suboptimal UX of GenAI tools for Programming}
GitHub Copilot proved useful in many ways, such as obtaining summaries of legacy code or unfamiliar codebases (see section 4.2 for more). For some, the \textit{``experience I have with GitHub Copilot is very good''} (P37) or \textit{``bearable''} (P32). 
However, a number of BLVSPs reported that GitHub Copilot provided \textit{``awful''} (P24) user experience (UX) with auditory overloads, despite the robust quality of the code it suggested. Unlike sighted users who can visually scan code suggestions and simply keep typing to ignore it, screen reader users receive Copilot's suggestions through audio. An audible cue (e.g., a beep) is followed by the screen reader reading the suggestion aloud. This adds yet another layer of auditory load on top of the screen reader’s continuous narration of the interface. Thus, its default settings  were \textit{``really, really, chatty... They're not sensible defaults to [BLVSPs] at all''} (P24): 
\begin{quote}
    \textit{``You're writing code and it's constantly giving you suggestions, but the default settings are for it to beep every time it gives you a suggestion, plus then it interrupts speech to tell you what the suggestion is.''} (P24)
\end{quote}

This constant aural communication, while intuitive for sighted professionals who could \textit{``easily go back and forth.. just click and go wherever''} (P13), significantly cut down on the productivity of BLVSPs and took them \textit{``out of the zone''} (P24). The speech interruption from Copilot extensions added another layer of aural information on top of screen reader outputs, making swift context switching challenging: \textit{``It's not easy, figuring out the best way [to] easily switch between that and talk to it in the same window with a Copilot extension''} (P13). P13 even preferred to pair program with a colleague for this very reason, comparing GenAI to an \textit{``unpaid intern... it does the job well. But relying on it fully is not a good idea.''}

To cope with these issues, our participants had to either \textit{``customize and disable certain things, or just train yourself to ignore a bunch of it''} (P13). Some felt that \textit{``there's no reason to install the [IDE] extension''} (P34), ignored it, and used web-interfaces to get answers \textit{``on-demand''} (P7) and \textit{``get the information when I want it, not when the AI thinks I want it''} (P7). P12 also \textit{``almost always''} used Copilot's web interface for the same reason.

\paragraph{Suboptimal UX of Other GenAI Tools}
Many GenAI tools (e.g., Claude, Gemini), despite their capabilities, were \textit{``not the most convenient interface to use''} (P7) for BLVSPs. ChatGPT was the tool that disappointed our participants the most, as \textit{``screen reader navigation is not really well thought-out''} (P15). In contrast to Gemini which provided relatively straightforward navigation \textit{``with headings for each answer''} (P6), many BLVSPs lamented the difficulty of navigating between responses without headings and the absence of button labels: \textit{``You always have to reread your own prompt before you get to the reply of ChatGPT''} (P10). P33 pointed out that difficulty of navigation skyrocketed when different elements were mixed together without proper distinction: \textit{``There's a code segment that you can indicate in HTML that code is code. But those gen AI tools don't indicate those things correctly.''}

Further compounding the challenges, frequent updates of GenAI tools many times broke accessibility. Buttons shifted back and forth between being labeled and unlabeled, and graphics were inconsistently accompanied by alternative text
, which forced users to constantly \textit{``re-learn how to use the tools with your assistive technology''} (P9) with every update. Additionally, OpenAI lacked robust technical support; P15 described that \textit{``they don't actually have a contact form for reporting issues with their website,''} and P7 shared, \textit{``It's really hard to actually get through to them.''} Disappointing navigation experiences, along with inconsistent updates breaking accessibility and the absence of proper feedback systems created a lose-lose situation for both the company and our participants: \textit{``I feel that they are missing out on their design principles... It's bad for me, it's bad for the company as well''} (P9). 

Distressed by the inaccessibility of these tools, BLVSPs sometimes chose to DIY their own scripts and applications \textit{``that are just using the API of [AI]''} (P13) so they \textit{``don't have to use their stupid interface''} (P16). Participants also felt left behind their sighted colleagues with new advances in GenAI tools (e.g., Cursor code editor) that has \textit{``so much potential''} (P7), but is not yet equipped with accessibility: \textit{``I've seen sighted developers who really know what they're doing crank out so much code with Cursor [code editor]. It's easily better than Copilot. Of course accessibility is not there. And so part of me does wonder from that perspective, ‘What am I missing?'''} (P7) Moreover, when blind people encounter inaccessibility in certain technologies, they may not go back: \textit{``Maybe they fixed it... But I can't really be bothered to try... The initial experience was so poor that I just have zero interest in going back and seeing if they've made any changes''} (P24).


\subsubsection{\textbf{Costs of Using GenAI}}
Though GenAI offered benefits like greater productivity and enhanced professional images, these also became risks for BLVSPs who were unable to quickly identify and fix errors as easily as their sighted colleagues.

\paragraph{Meticulous Validation}
While our participants acknowledged that \textit{``there's a magical quality to [GenAI]''} (P16), persistent hallucinations bred frustration. Frequent hallucinations led our participants to liken GenAI to \textit{``a very arrogant piece of metal... its opinion about its work is excellent''} (P10), and compared their interactions with GenAI to \textit{``talking to an overly confident drunken acquaintance''} (P26), highlighting the need to \textit{``
always, always... laterally research''} (P26). 

As opposed to sighted professionals, for BLVSPs detecting hallucinations was \textit{``really taxing''} (P14), as it took many tabs with a screen reader, or sometimes even impossible (e.g., descriptions of technical visuals). When hallucinations persisted, participants had to \textit{``go and do extra work to fix it''} (P39), which \textit{``somewhat defeats the purpose''} (P27): \textit{``You struggle hard to understand what it did and to fix it, and end up spending much more time than if you took the other path and just wrote the damn thing yourself''} (P10).

\paragraph{Increased Vulnerability}
BLVSPs' perceived that their use of GenAI tools carried higher risks than their sighted colleagues'. They felt that \textit{``as blind people we stand to benefit the most from them''} (P24), but at the same time felt uniquely more \textit{``vulnerable''} (P13) to hallucinations because they were \textit{``the group that's probably least able to tell whether they're hallucinating or not''} (P24).  Consequently, BLVSPs still had to verify GenAI outputs with their colleagues, even though one of their main reasons for turning to GenAI was to avoid repeatedly asking for sighted assistance, which could be detrimental to their professional reputations. P7 described:
\begin{quote}
    \textit{``Without asking another human and being the annoying colleague who's like, ‘Hey, can you look at this?',  there's absolutely no way for me to know if it is hallucinating. And I think that's one of the things that does result in an unequal playing field.''} (P7)
\end{quote}

In addition, some believed that they used GenAI for certain \textit{``riskier'' }(P24) tasks that their colleagues wouldn't use it for, such as analyzing log files when debugging. While participants reported that GenAI tools greatly improved the accessibility of debugging activities which are known to be inaccessible due to the amount of noise in logs, P24 \textit{``worr[ied] that sighted people aren't going to be using [GenAI] to analyze the logs''} and an unspotted error could make make him \textit{``look stupid.''} In addition, P35 worried about inconsistent formatting of GenAI-generated content when combined with his own work, fearing he \textit{``might miss something and... actually ship that out [to colleagues],''} potentially impacting how their work appeared to others if even a small portion looked \textit{``different in terms of fonts and stuff like that.''}

P7 articulated the difficult position BLVSPs are stuck in: while GenAI can't always be trusted, they \textit{``have to trust it, that what it's saying is actually true''} to \textit{``get the work done''} quicker and avoid constantly asking for sighted assistance. This reliance, coupled with the lack of \textit{``safety net''} options (P35) available to our participants amplified the risks that BLVSPs took in using GenAI for work: \textit{``they [sighted people] have so many options to fall back on, like tutorial videos and stuff which is, again, much harder for blind folks. And it could many times be a case where I am either learning [programming] through this [AI] or I am not learning it at all or something like that. So, either generating code or not generating at all''} (P35).

\subsubsection{\textbf{Non-Use vs. Deciding to Take the Risk}}
Faced with unreliable GenAI outputs and higher risks compared to their colleagues, some users deliberately chose non-use in professional settings and relied on existing workarounds or developed DIY tools, BLVSPs carefully contemplated whether to find another option or still proceed with using GenAI for completing the task at hand. Some participants \textit{``ask[ed] a couple questions''} (P7) to themselves on whether GenAI would be a suitable solution. P38 noted developing an intuition: \textit{``I know what kind of questions I should be asking and what kind of questions I shouldn't be asking,''} recognizing GenAI's \textit{``limited capacity to remember and think.''} As such, they used GenAI in \textit{``low-stakes''} tasks such as generating boilerplate code and brainstorming possible ideas and solutions, rather than tasks with high complexity or those that required definitiveness 
While some BLVSPs chose to abandon GenAI tools and opted for already established workarounds, others continued using GenAI tools when \textit{``clearly the reward outweighs the risk''} (P24).

\subsection{GenAI's Wider Impact on Workplace Equity} \label{44}
Many participants believed GenAI was a \textit{``game-changer''} (P5, P17). GenAI \textit{``open[s] a huge space of possibilities... It is extremely promising to make the lives of people with disabilities more independent, and more equal with people without disabilities''} (P21). 
However, they simultaneously believed GenAI had the potential to set them further back if such tools remain inaccessible or not allowed for use at work. 

\subsubsection{\textbf{GenAI and Accessibility (Awareness) in the Workplace}}
GenAI supported general workplace accessibility, but at the same time raised concerns about how general awareness about the importance of accessibility among software professionals would decrease, which could potentially perpetuate inaccessible developer tools. 


Participants expressed concerns of a potential \textit{``scary trend''} (P32) that those without much experience in accessibility will assume GenAI to be the \textit{``magic powder''} (P15) that solves all accessibility problems: \textit{``People just believe that AI will do it somehow, and they're pretty cool with that''} (P15). Many worried that developers will entirely trust GenAI to take care of accessibility, thereby losing motivation, and absolving them of the need to attend to accessibility in tools they create:
\begin{quote}
    \textit{``They just won't bother including best accessibility practices... They won't bother with writing descriptions, labeling buttons properly. They'll just believe that they won't have to pay the price for making this accessible and they will just externalize [to GenAI] the cost of using their product.''} (P15)
\end{quote}
Participants expressed frustrations about GenAI being used as an alternative for human professionals with accessibility knowledge to \textit{``cut some costs"} (P15), as currently humans were far more superior: \textit{``It's not the same. I'm sorry, soy milk is not the same as dairy milk''} (P32). 
P35 pointed out that when GenAI produces inaccessible code and software, \textit{"it [would be] very... hard to fix''} and pinpoint the accessibility problem since GenAI took over so much of the process. 

Notably, one participant thought otherwise; P10 believed that accessibility being GenAI's entire responsibility may be better, as the average developer's knowledge about accessibility in software products is very poor: \textit{``I think that [Gen]AI knows more accessibility than other [sighted] developers... Because many developers don't know anything at all about accessibility... I saw enough inaccessible things [built by sighted developers].''} (P10)

\subsubsection{\textbf{GenAI's Impact on Perceived Career Outlook}}
Although our participants echoed the concern that \textit{``GenAI gets way better than it's now and starts replacing developers''} (P10), they also felt that GenAI would make software engineering training materials more accessible, giving them access to a wider range of job positions, and potentially leading to recognition from superiors.
For example, P37 recounted watching YouTube tutorials that used ample ç language when learning programming: \textit{``They were not very accessible...: `Click on this. Go to that. Go on that and do this and do that'... I was not getting it... I struggled a lot while learning programming.''} However, GenAI provided an easier entry point to coding: \textit{``You can ask [Gen]AI to list down the concepts in simpler language or in the way you understand the most.''}
P31 also shared a story of his acquaintance who used GenAI tools to learn programming, and that \textit{``he was more confident in wanting to go into the tech industry to do tech-type work.''} As such, P38 believed that \textit{``more visually impaired people will be participating in the IT industry because of GenAI... people would be using it to learn programming a lot.''}

Further, participants envisioned GenAI opening up more career opportunities and positions (e.g., UI/UX designer, prompt engineer), some of which were previously considered unattainable for BLVSPs because of the visual elements: 
\begin{quote}
   \textit{``Could you be a UX designer if you were blind, using GenAI? ... Maybe they wouldn't be able to use Photoshop to design a button, but maybe if they understood what they wanted, they'd have enough sight to feed that into a model and then make sure that it made the right thing.''} (P24)
\end{quote}
Some even saw \textit{``more streams opening up''} (P9) for professionals with disabilities in the technical track, instead of pigeonholing them into accessibility roles against their interests.

Several participants believed the time efficiencies gained by GenAI would enable them to produce faster, higher-quality deliverables that make them \textit{``more likely....to get noticed''} and \textit{``help me climb the ladder''} (P7). Not only would this advance their careers, but it would help them demonstrate to superiors that they are as productive and \textit{``also equally [as] capable as a sighted person''} (P5), thus contributing towards their career success.

\subsubsection{\textbf{Organizational Policies around GenAI Use}}
Our participants' organizations had different policies around using GenAI in the workplace, which impacted them in various ways. 

\paragraph{Permission and Encouragement}
Many companies either encouraged the use of GenAI at work or did not restrict it, as long as no security or privacy was breached. P24 reported that their organization changed policies into allowing select models for work, and shared that \textit{``it was a challenge resisting the urge to use them when I knew how helpful they could be.''} Notably, some companies even built internal LLMs for more convenience, but P10 was blocked from using it due to its inaccessibility: \textit{``As a company policy, they actually built a local ChatGPT for us to use, with some customizations. I'm not sure what it does, because it's inaccessible.''}

\paragraph{Limited Use, Selective Disclosure}
Four participants' companies only allowed select models that passed internal checking. One participant who experienced remarkable benefits from GenAI, in fear that going too much into detail about their specific use case would result in restrictions around their GenAI use, kept their use case \textit{``quite generic''} and only selectively disclosed how they used GenAI to their organization: \textit{``I'm saying, `...I'm going to be writing some code that uses these models. It's not audience facing. It's just for me.' And I don't think that was a lie, but it was a beginner-friendly version of what I was going to be doing with them.''}

\paragraph{Total Ban, Surreptitious Use}
Our participants\footnote{Here, we omit all participant IDs due to risks of identification.} speculated \textit{``quite a bit of unauthorized use of models that were happening internally''} in companies where GenAI use was prohibited altogether. Five out of eight participants whose companies disallowed use disclosed clandestine use.  Even those who were currently allowed to use GenAI emphasized that \textit{``I won't work at a company that doesn't use GenAI''} or further, \textit{``would happily use it under the radar.''} 
BLVSPs described the tension that they experienced as they were not willing to give up the productivity boost granted by GenAI, especially as a BLVSP: \textit{``Do I use it behind the scenes and not tell it and keep my job? Or do I just not use it and risk lower productivity and being slower?''} Three participants cleverly \textit{``use[d] it anyway on my personal laptop''} while not using it on corporate devices to \textit{``save myself''}.

Participants were also \textit{``very alert to not leave AI-generated materials that get me in trouble.''} Sometimes, they \textit{``d[id]n't have a choice''}: \textit{``I will try my best to not use it... unless I'm absolutely forced into it... [If] they have something inaccessible...''} Another BLVSP further explained the rationale behind such decisions: \textit{``I wouldn't even say it's malicious intent... at the end of the day... I can't be pressing a bunch of buttons, to go back to the window, to see the message that someone sent in the chat while people are waiting for me [in a meeting].''} Interestingly, one, whom other employees occasionally report to, shared that they have \textit{``turned a blind eye to it [GenAI use] a few times, and will probably continue to do so,''} as long as \textit{``I see that they're using AI responsibly and... their work does seem to be better in certain ways.''} 

\section{Discussion}


Existing scholarship focuses on how software developers use GenAI to code~\cite{davilaIndustryCaseStudy2024, bird_copilot, flores_impact_2025} and how BLV individuals use GenAI for accessibility~\cite{adnin_blind_genai_2024, tang_everyday_uncertainty}. To the best of our knowledge, this is the first research study to investigate how BLV professionals who work in software development utilize GenAI for everyday job tasks, including but not limited to writing code at the workplace.

\subsection{To Use or Not to Use? The Higher-Risk, Higher-Return Calculus of GenAI }

The \textit{``game-changing"} returns of using GenAI were evident in several ways: beyond its support for programming, GenAI enabled independent engagement in highly visual job tasks (e.g., debugging, UI verification, interpreting technical diagrams) that previously demanded either significant manual effort (e.g., DIY tooling~\cite{cha_dilemma_2025}) or routine reliance on sighted colleagues~\cite{das_ideation_24}. BLVSPs were also able to avoid unwanted disability disclosure as well as excessive, unpaid work dealing with inaccessible tools and tasks, which participants reported enhanced their mental well-being. These benefits in turn had a positive impact on their outlook for their future careers, as they were able to take on more responsibilities and diverse roles 
and be perceived by their colleagues as more competent, productive, and self-sufficient. 
GenAI also expanded their problem-solving toolbox, perhaps more so than it did for their sighted counterparts. 

However, accompanying these substantial benefits were substantial risks, often exceeding those faced by BLVSPs' sighted counterparts. A primary source of risk was the suboptimal accessibility and user experience of GenAI tools themselves. Our work confirms previously documented factors that inhibit the usability of GenAI tools (e.g., inaccessible UI~\cite{adnin_blind_genai_2024}, cognitive overload~\cite{flores_impact_2025}). We document additional factors (e.g., inability to distinguish code and URL output from plain text, frequent updates breaking accessibility) as well as situating them within the workplace context. 
These accessibility and usability challenges forced BLVSPs to perform \say{access labor} in re-learning how to use a given tool, building DIY tools to avoid these issues altogether, or finding other, manual workarounds, all of which their sighted colleagues did not experience. This led to lower productivity and decreased well-being, in some cases negating the gains of GenAI. 

Further compounding this risk was what we refer to as the \say{verification catch-22.} While hallucinations were a general concern among both sighted~\cite{pereira2025exploring} and blind software professionals, BLVSPs reported feeling uniquely more vulnerable.
Sighted software professionals have multiple ways to rapidly confirm an answer (e.g., eyeballing the response, cross-checking with documentation, browsing Stack Overflow). However, because of the linear, aural information presentation of screen readers, BLVSPs are unable to quickly scan results as well as quickly switch back-and-forth between tools and documents. Further, referential sources (official documentation) are not always fully accessible~\cite{storer_Infoseeking_21}. As such, detecting errors was highly demanding or impossible, which amplified the risks BLSVPs took, and threatened to nullify the returns of GenAI tools when verification labor took too long or was unsuccessful. This created a significant irony: one of the main reasons BLVSPs chose to use GenAI was to avoid asking for sighted assistance to maintain a professional image, yet using GenAI meant they frequently had to either consult a sighted colleague or trust unverified output, thus risking their professional image.



Based on the higher-risks and higher-returns of GenAI tools identified, our participants carefully contemplated the decision to adopt or forgo GenAI on a case-by-case basis. On one hand, \textit{using GenAI} helped them compete with sighted colleagues for recognition and promotions, but it came with increased risks of lost productivity and credibility, and threats to their job security if caught with unauthorized use. On the other hand, \textit{not using GenAI} carried the potential to widen the gap between them and their sighted colleagues and hinder their careers, as tasks would still remain inaccessible, additional labor to compensate for accessibility was needed, and sighted colleagues would be moving forward with their own use of GenAI tools. In navigating these tradeoffs, BLVSPs often chose to bear the risk in pursuit of the benefits GenAI brought them.

Furthermore, while GenAI had the potential to reduce the gap with their colleagues, we note that the \say{equal playing field} may be a moving target, as sighted professionals are also rapidly adopting and leveraging GenAI to further their own productivity and capabilities (e.g., ~\cite{ziegler2024measuring, khojah_beyond_2024}). BLVSPs thus might find themselves in a continuous race to catch up, especially in a world where demand for skills and jobs related to AI are expected to see rapid growth~\cite{USBLS_2025, WEF_Future_Jobs_2025}. 


\subsection{Towards a Low-Risk, Higher-Return GenAI}

GenAI tool \textit{developers} must consider inclusive design to support non-visual access, such as making sure that GenAI tools comply with digital accessibility guidelines (e.g., Web Content Accessibility Guidelines (WCAG)~\cite{WCAG}). One key principle in the WCAG is ``perceivability"---``information and user interface components must be presentable to users in ways they can perceive"~\cite{WCAG}--- and this could be achieved through ensuring seamless screen reader compatibility and navigation, non-overwhelming auditory cues~\cite{flores_impact_2025}, and accurate labeling of all elements (e.g., code, URLs) so they are distinguishable. Also included in this would be consistency and stability with updates, to prevent regressions in accessibility that keep BLVSPs trapped in a cycle of constant re-learning.

\textit{Researchers} should study and formulate best practices in building accessible GenAI tools, so tool developers can use these guidelines to inform their work. This requires longitudinal studies investigating persistent and newly emerging effects of GenAI tools on BLVSPs (e.g., productivity, well-being, career progression) and the design of robust, non-visual verification methods to support BLVSPs in detecting hallucinations. Feasible ideas include providing internal self-validation methods~\cite{adnin_blind_genai_2024}, high-quality alternative texts for visual descriptions, and transparent confidence metrics (e.g., percentages), and third-party remote verification options. 

Finally, we urge \textit{employers} to develop clear and inclusive GenAI use policies that recognize the needs of BLVSPs. Rather than imposing uniform restrictions, policies should consider each BLVSP's needs and allow GenAI to be an accommodation with flexibility. In addition, procurement departments should ensure GenAI tools that are brought in for employee use---either commercially available or internally built GenAI tools---are accessible so BLVSPs can access and utilize them effectively and efficiently. Organizations should be cognizant that GenAI tools are not a panacea and will not address all issues; therefore accessibility awareness must be retained and inclusive work environments should continue to be promoted.

\section{Limitations and Threats to Validity}
In this section we discuss the credibility, reliability, and transferability of our findings as is common in qualitative research~\cite{creswellQualitativeInquiryResearch2018}.


\paragraph{Credibility.} The research team consists of scholars in multiple fields:expertise in accessibility research, working with people with disabilities, SE research experience, including one researcher with extensive industry experience. The team thus brought different perspectives to the research project. Such diversity helped the study in several ways such as minimizing risks arising from individual researchers' biases, ensuring the questions in the interview protocol were relevant to software professionals who were BLV, and interpreting the interview data through the lenses of accessibility and software engineering.



\paragraph{Reliability.} To address risks related to inconsistencies in capturing and analyzing data
, we ensured consistent data collection and analysis processes, as detailed in \autoref{study_design}. We provide the interview protocols and participant information for inspection in supplementary materials~\cite{Dryad_SuppData}.

While we did not triangulate with additional data sources, we note that some of the identified themes (e.g., Copilot accessibility problems, using AI for image descriptions) resonate with and extend prior research into the GenAI experiences of BLV people more broadly (e.g., \cite{flores_impact_2025, adnin_blind_genai_2024}), although not in specific workplace settings, thereby increasing confidence in the accuracy of these findings. 

We note that we did not always probe which specific GenAI tools or models participants used, thus parts of our findings do not specify the exact tool referenced.

\paragraph{Transferability.} Our participants held a wide range of job roles and tenures, and were employed in various industry sectors and across different company sizes (see \autoref{study_design}). Our sample had a gender skew towards men (see \autoref{sec:demographics}), which is not uncommon in studies of software professionals given the overall demographic.
The majority of participants resided in the USA, with other regions represented, including Europe, Asia, and South America. Thus, our findings do not represent all countries and cultures. 

Finally, a particular threat to acknowledge is that new LLMs and GenAI-based tools are being released frequently; therefore we acknowledge that noted accessibility issues with specific GenAI tools may have been addressed already. If any of the noted issues have been addressed, this illustrates the observation that the accessibility of tools is often considered an afterthought~\cite{wentz_retrofitting_2011}, leaving BLVSPs unable to take immediate advantage of novel, game-changing tools when they are first made available. 
\section{Conclusion}
GenAI tools such as ChatGPT and GitHub Copilot are rapidly becoming an integral part of software development. Yet, little is known about their impact on the workplace experiences of minority groups, such as blind and low vision software professionals (BLVSPs). This qualitative study with 39 BLVSPs explored the benefits and risks of using GenAI and its impact on career prospects. We found that GenAI was seen as a game-changer for BLVSPs, anticipation of making significant progress in leveling the playing field with sighted software professionals (for reasons such as increased productivity and reduced needs for workarounds). However, we also found that BLVSPs faced risks in using GenAI, beyond those faced by sighted developers (e.g., being more vulnerable to hallucinations, struggling with poor UX, leaving poor impressions on colleagues). The thorny choice to use or not to use GenAI posed outsized tradeoffs to BLVSPs, leaving them more vulnerable to reputational damage and career setbacks.
Recommendations include GenAI tool developers to adhere to digital accessibility guidelines and for researchers to investigate robust non-visual verification systems.

\begin{acks}
We would like to thank our participants for their involvement in our study and for providing valuable insights and perspectives. This work was supported by the National Science Foundation (NSF) awards \#2211790, \#2326023, \#2210812, and \#2326489. 
\end{acks}

\bibliographystyle{ACM-Reference-Format}
\bibliography{manuscript}

\end{document}